\documentclass[twoside]{dis04}
\usepackage{epsfig}
\usepackage{amssymb}
\def\runauthor{M. Deile}
\def\shorttitle{TOTEM: FORWARD PHYSICS AT THE LHC}
\begin{document}

\title{TOTEM: FORWARD PHYSICS AT THE LHC}

\author{Mario Deile\\
on behalf of the TOTEM collaboration}

\address{CERN\\
1211 Gen\`{e}ve 23, Switzerland
}

\maketitle

\abstracts{
The TOTEM experiment with its detectors in the forward region of CMS
and the Roman Pots along the beam line
will determine the total pp cross-section via the optical theorem
by measuring both the elastic cross-section and the total inelastic rate.
TOTEM will have dedicated runs with special high-$\beta^{*}$ beam optics
and a reduced number of proton bunches resulting in 
a low effective luminosity between $1.6 \times 10^{28} {\rm cm}^{-2} 
{\rm s}^{-1}$ and $2.4 \times 10^{29} {\rm cm}^{-2} {\rm s}^{-1}$.
In these special conditions also an absolute luminosity measurement will 
be made, allowing the calibration of the CMS luminosity monitors needed at 
higher luminosities.
The acceptance of more than 90\,\% of all leading protons in the Roman Pot
system, together with CMS's central and TOTEM's forward detectors extending to 
a maximum rapidity of 6.5,
makes the combined CMS+TOTEM experiment a unique instrument for exploring
diffractive processes.
Scenarios for running at higher luminosities necessary for hard diffractive
phenomena with low cross-sections are under study.
}

\section{The TOTEM Experiment: Layout and Detector Technologies} 
\begin{figure}[!h]
\begin{center}
\epsfig{file=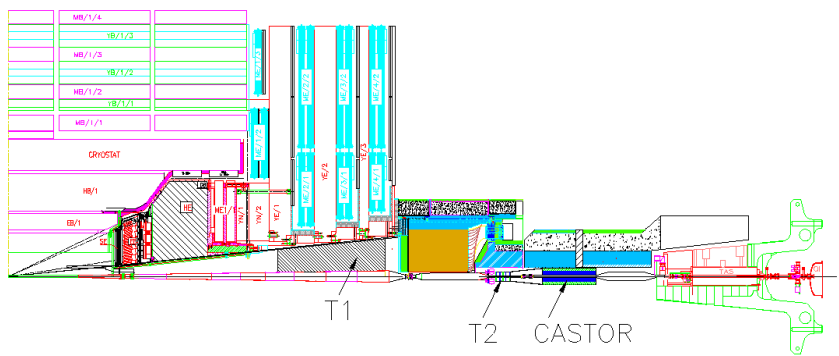,height=5cm}
\caption[*]{The CMS detector with the TOTEM forward trackers T1 and T2.
Note also the planned forward calorimeter CASTOR (under CMS's responsibility).}
\label{fig_CMSlayout}
\end{center}
\end{figure}

The TOTEM experiment~\cite{TDR} situated at IP5 of the LHC, complements the 
CMS detector by the forward trackers T1 and T2 inside CMS
(Figure~\ref{fig_CMSlayout}) 
and by 
a system of Roman Pot stations at distances of 147\,m, 180\,m and 220\,m
from the interaction point. 

The T1 telescopes on both sides of the interaction point will consist of 
five planes of Cathode Strip Chambers (CSC) covering the pseudorapidity range 
$3.1 < |\eta| < 4.7$. Each detector will measure three projections, rotated
by 60$^{\rm o}$ with respect to each other: the two sets of cathode strips
with a pitch of 5\,mm will have an analogue read-out whereas the digital 
read-out of the anode wires with a pitch of 3\,mm will provide trigger
functionality.

For T2, extending the acceptance into the range $5.3 < |\eta| < 6.5$, the
Gas Electron Multiplier (GEM) technology as used successfully in 
COMPASS~\cite{compassgem} has been chosen. The read-out plane will have
both pads ($\Delta \eta \times \Delta \phi = 0.06 \times 0.017 \pi$) 
connected to digital VFAT chips for triggering, 
and circular strips with 400\,$\mu$m pitch read out with the analogue APV25.

The Roman Pot detector system is optimised in view of measuring proton 
scattering angles down to a few $\mu$rad. The detector edges have to approach
the beam to a distance of $10\, \sigma_{beam} + 0.5$\,mm $\approx$ 1.3\,mm.
In order to minimise the dead space near the detector edge, TOTEM develops
Silicon detectors with full 
efficiency up to less than 50\,$\mu$m from their physical edge. Three approaches
are under investigation: (1) the novel 3D detector technology with 
pillar-shaped electrodes processed through the Silicon bulk and active edges; 
(2) planar Silicon
detectors with active n$^{+}$ doped edges; (3) planar Silicon detectors with
a very narrow ($\sim 50\,\mu$m) current-terminating guard-ring structure.
The electrode pitch is 66\,$\mu$m resulting in a spatial resolution of 
about 20\,$\mu$m per plane.
For all three technologies first prototypes have been built and tested
with X-rays or muon beams. The width of the efficiency transition from 
10\,\% to 90\,\% was found to range between 15\,$\mu$m (1) and 50\,$\mu$m (3).
Further muon beam tests, for the first time with final-sized detectors 
($\sim 9\,{\rm cm}^{2}$) and close-to-final electronics will be performed in 
summer / autumn 2004.

\begin{figure}[!h]
\begin{center}
\mbox{
\epsfig{file=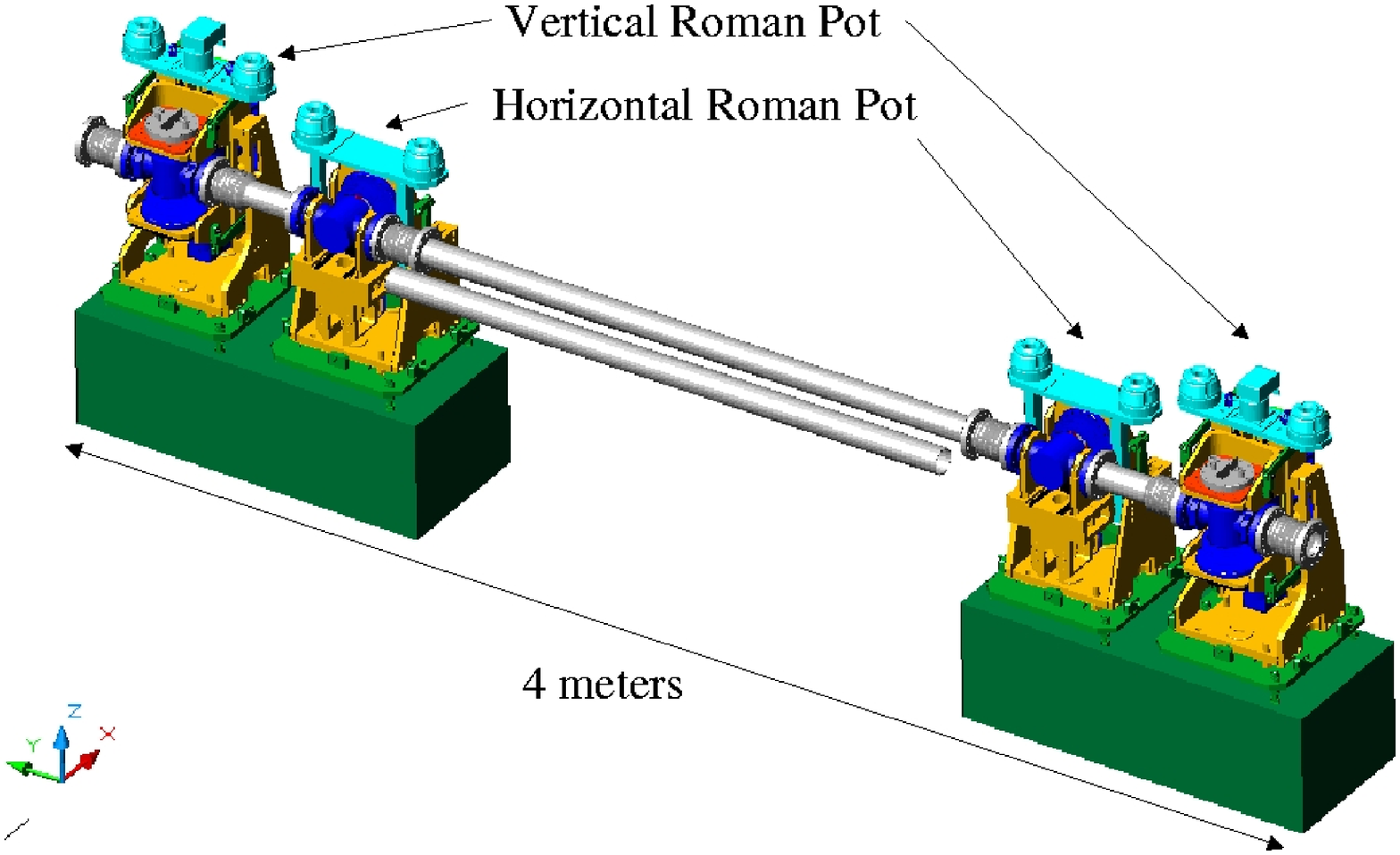,height=5cm}
\epsfig{file=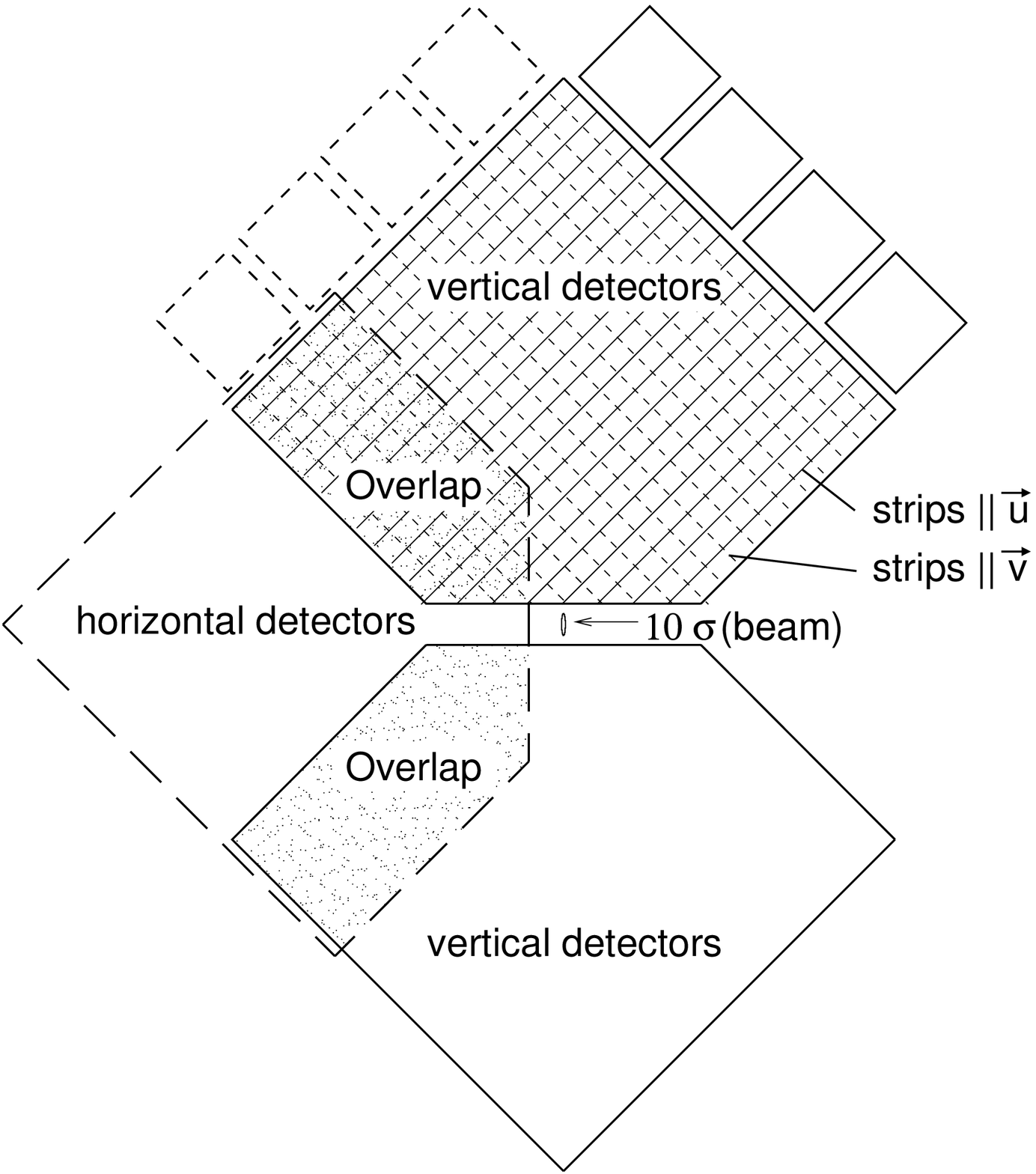,height=5.5cm}
}
\caption[*]{Left: Roman Pot station. Right: arrangement of the detectors in 
the two vertical and the one horizontal Roman Pots of a station.}
\label{fig_overlap}
\end{center}
\end{figure}
Each Roman Pot station (Figure~\ref{fig_overlap}a) consists of two units
with a distance of 4\,m. Each unit has two vertical pots approaching the
beam from the top and the bottom, and one lateral pot sensitive to 
diffractive protons. 
Furthermore, the overlap
between the horizontal and the vertical pots (Fig.~\ref{fig_overlap}b) will 
serve for measuring the relative distance of the vertical detectors.
Each pot will contain 10 planes of Silicon detectors, 6 equipped with 
analogue APV25 read-out chips for precise tracking, and 4 equipped with 
digital VFAT chips providing trigger and tracking information.

\section{Beam Optics and Running Scenarios}

For the precise measurement of proton scattering angles 
on the few $\mu$rad level
a special beam optics scheme with $\beta^{*} = 1540\,$m was developed.
It is characterised by a small beam divergence in the interaction point 
(0.29\,$\mu$rad) and a focal point in the Roman Pot station at 220\,m for
both the horizontal ($x$) and the vertical ($y$) track projections.
In order to avoid parasitical bunch crossings downstream of the nominal
interaction point due to the parallelism and large width ($\sim 0.4$\,mm) 
of the two beams, the number of 
bunches will be reduced from 2808 to initially 43 and later 156. 
Scenario 1 in Table~\ref{tab_scenarios} will serve for measuring the elastic
cross-section at low $|t|$, the total cross-section, the absolute 
luminosity and soft diffraction. To minimize the emittance, these runs
will be done with a smaller bunch population.
Higher luminosities for (semi-) hard diffraction (scenarios 2 and 3) 
can be reached by increasing the number of bunches and the bunch
population. The details of the optics for Scenario 3 are still under 
development. Scenario 4 was introduced for measuring elastic
scattering at large $|t|$.

\begin{table}[ht]  
\begin{center}  
\scalebox{0.95}{
\begin{tabular}{|l|c|c|c|c|c|}\hline 
Running scenario & 1 & \multicolumn{2}{|c|} {2} & 3 & 4 \\
\hline
 $ \beta^*$ [m] & 1540 & \multicolumn{2}{|c|}{1540} & 200--400 & 18 \\ 
\hline
Number of bunches & 43 & \multicolumn{2}{|c|}{156} & 936 & 2808 \\ 
\hline
Protons per bunch & $0.3 \cdot 10^{11}$ & 
$0.6 \cdot 10^{11}$ & $1.15 \cdot 10^{11}$ & $1.15 \cdot 10^{11}$ & $1.15 \cdot 10^{11}$  \\ 
\hline
Transverse norm. & 1 & 1 & 3.75  & 3.75 & 3.75 \\  
emittance [$\mu$m rad] & & & & & \\ 
\hline
beam size at IP [$\mu$m] & 454 & 454 & 880 & 317--448& 95 \\ 
\hline
beam divergence & 0.29 & 0.29 & 0.57 & 1.6--1.1& 5.28 \\ 
at IP [$\mu$rad] & & & & & \\
\hline
$\frac{1}{2}$ crossing angle [$\mu$rad] & 0 & \multicolumn{2}{|c|}{0} & 100--200 & 160\\
\hline
$\mathcal{L}$ [cm$^{-2}$s$^{-1}$] & 
$1.6 \cdot 10^{28}$ & \multicolumn{2}{|c|}{$2.4 \cdot 10^{29}$} &
$\sim 10^{31}$ & $3.6 \cdot 10^{32}$  \\ 
\hline
\end{tabular}
}
\caption{TOTEM running scenarios.}
\label{tab_scenarios}  
\end{center}  
\end{table}

While elastic scattering, the total cross-section and soft diffractive 
processes 
can be measured by TOTEM alone, hard diffraction will be studied together
with CMS, i.e. TOTEM will technically act as a subdetector with level-1 
trigger capability.
The trigger signal from the Roman Pots at 
220\,m arrive at the CMS global trigger still within its latency time.
With Roman Pots further away, level-1 triggering would not be possible.

TOTEM operation with normal LHC beam optics ($\beta^{*} = 0.5\,$m) and 
$\mathcal{L} \ge 10^{33}$\,cm$^{-2}$\,s$^{-1}$ is under study
but not yet part of the official TOTEM programme. 

\section{Physics Programme and Performance}

\subsection{Elastic Scattering}
\begin{figure}[!h]
\begin{center}
\mbox{
\epsfig{file=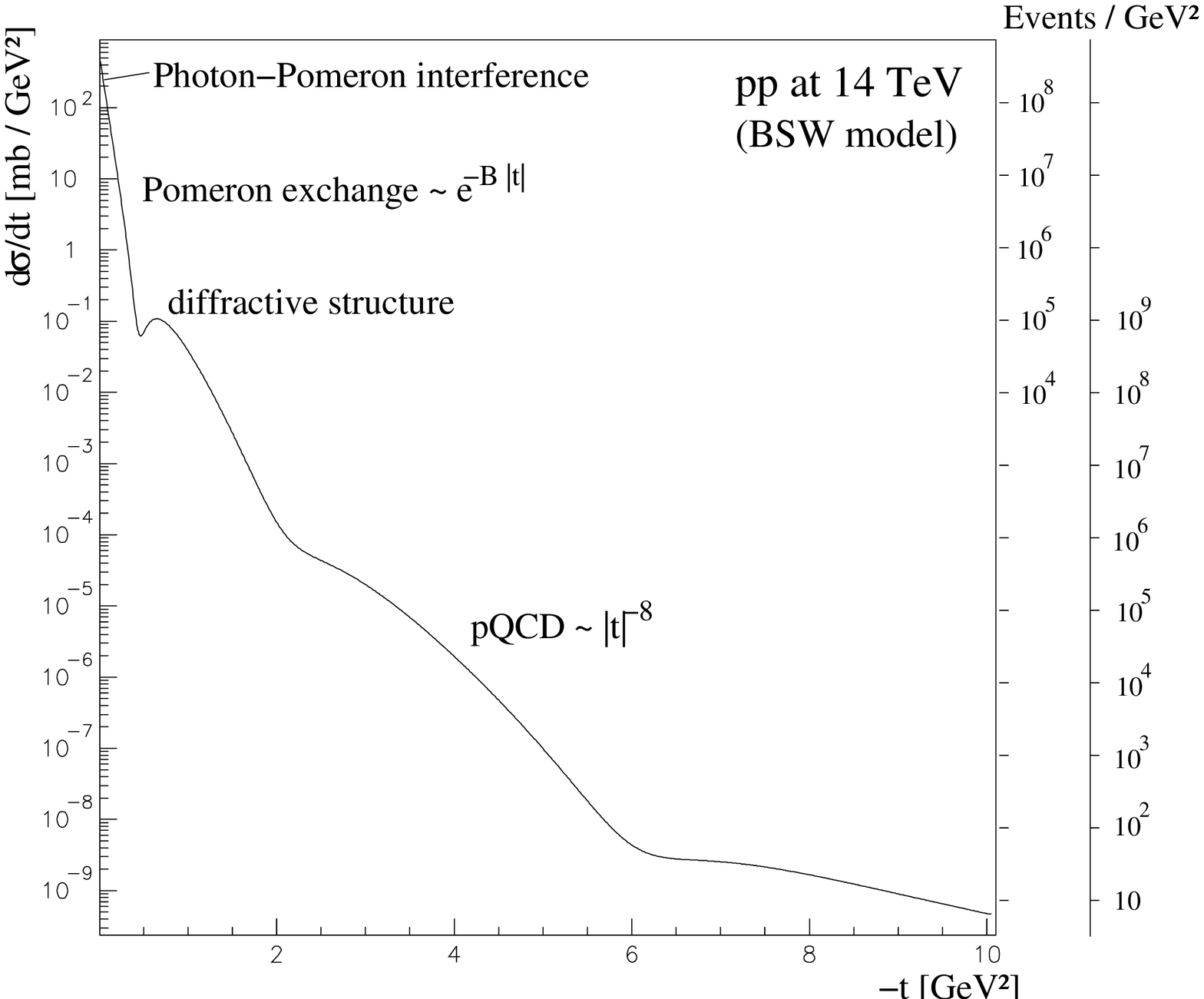,height=5.5cm}
\epsfig{file=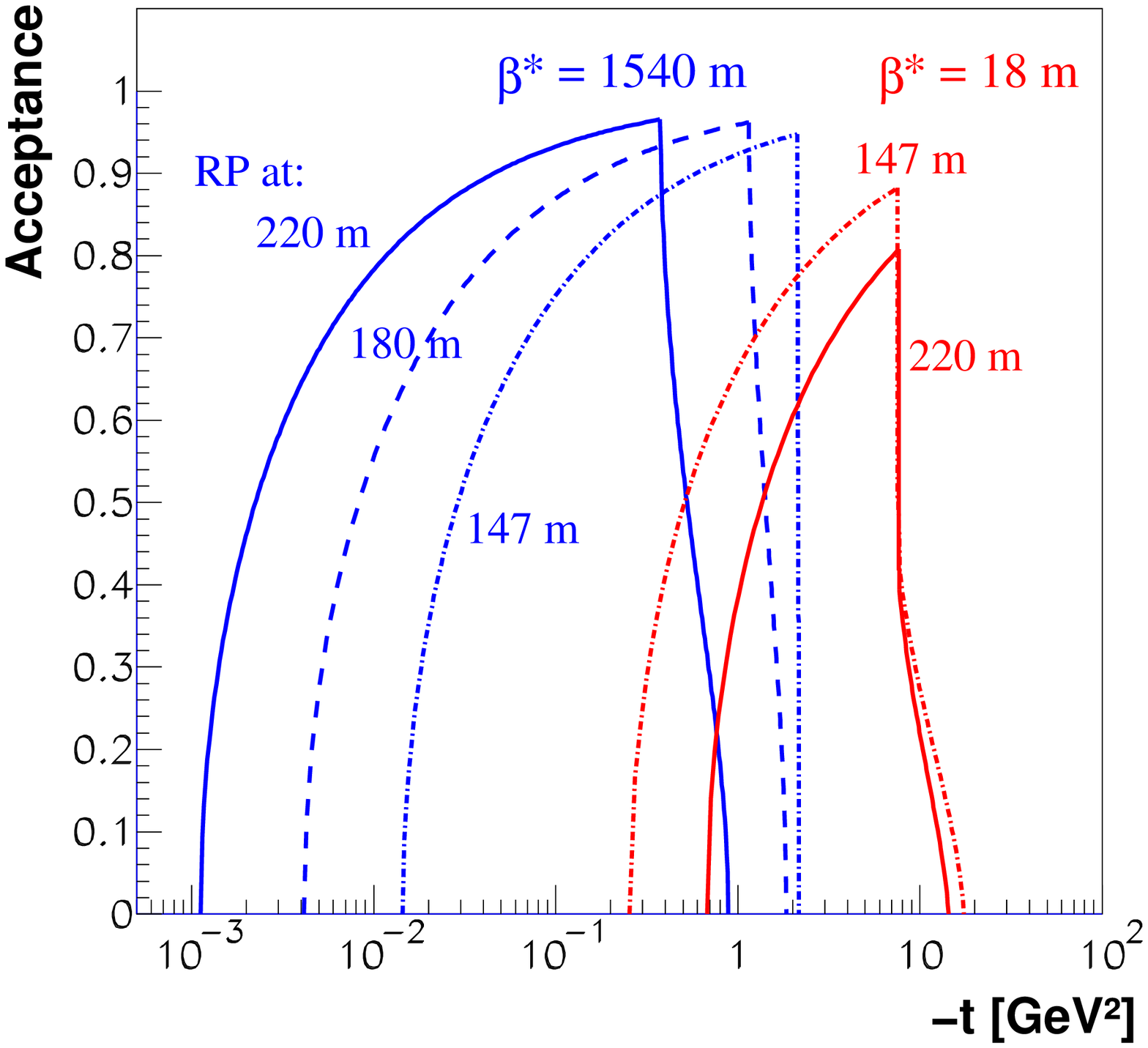,height=5.5cm}
}
\caption[*]{
Left: elastic cross-section predicted by the BSW model~\cite{bsw};
the number of events on the right-hand scales correspond to integrated
luminosities of $10^{33}$ and $10^{37}\,$cm$^{-2}$ (about 1 day with Scenarios
1 and 4). Right: 
t-Acceptance for elastic scattering with the 1540\,m and the 18\,m
optics and for the different Roman Pot stations.
}
\label{fig_taccept}
\end{center}
\end{figure}

TOTEM will measure the elastic pp cross-section in a $|t|$-range from 
$2 \times 10^{-3}\,$GeV$^{2}$ to 8\,GeV$^{2}$ (Figure~3, left). 
Since the acceptance for the $\beta^{*} = 1540$\,m optics ends at 2\,GeV$^{2}$
(Figure~\ref{fig_taccept}, right) and since at higher $|t|$ more luminosity 
is needed, the $\beta^{*} = 18$\,m optics was created to extend the 
range. Accessing the Coulomb and interference region 
at $|t| < 2 \times 10^{-3}\,$GeV$^{2}$ will be attempted either by approaching 
the Roman Pot detectors closer than $10\, \sigma + 0.5$\,mm to the beam or
by running the LHC with a reduced centre-of-mass energy $\le 6\,$TeV. 

The t-resolution achieved with the conditions of Scenario~1 and a detector
resolution of 20\,$\mu$m is 
$\sigma(t)/t \approx 7\,{\rm \%} / \sqrt{t/0.002\,{\rm GeV}^{2}}$.
Due to the 
parallel-to-point focussing at 220\,m in both track projections, a good
azimuthal resolution of $\sigma(\phi) \approx 50\,{\rm mrad} / \sqrt{t/0.002\,{\rm GeV}^{2}}$
is achieved, which 
helps reducing beam halo background by testing the collinearity of protons
detected in the two opposite arms of the experiment. Furthermore, the 
measurement of $\phi$ can help determining the parity of exclusively 
produced particles.

\subsection{Total pp Cross-Section and Luminosity}
The total pp cross-section and the luminosity will be measured via the 
Optical Theorem and are
given by
\begin{equation}
\sigma_{tot} = \frac{16 \pi}{1 + \rho^{2}} \cdot
\frac{dN_{el}/dt |_{t=0}}{N_{el} + N_{inel}}\:,
\qquad
\mathcal{L} = \frac{1 + \rho^{2}}{16 \pi} \cdot 
\frac{(N_{el} + N_{inel})^{2}}{dN_{el}/dt |_{t=0}} 
\end{equation}
The extrapolation of the elastic cross-section to $t=0$ suffers mainly from
insufficient knowledge of the functional form and from uncertainties in beam energy,
detector alignment and crossing angle. The expected systematic error 
is about 0.5\,\% whereas the statistical error is less than 0.1\,\% for only
10 hours of running. The uncertainty of the total rate ($N_{el} + N_{inel}$) is
given by trigger losses and beam-gas background; it amounts to 0.8\,\%.
Taking also into account the uncertainty in $\rho = 0.12 \pm 0.02$, 
$\sigma_{tot}$ and the luminosity can be measured with about 1\,\% precision.

\subsection{Diffraction}
\begin{figure}[h!]
\begin{center}
\epsfig{file=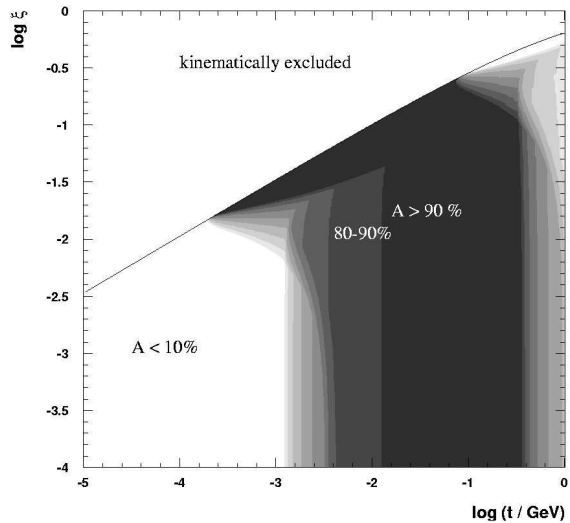,height=7cm}
\caption[*]{
Acceptance for diffractive protons with the 1540\,m optics 
in the RP station at 220\,m. 
}
\label{fig_txiaccept}
\end{center}
\end{figure}

The acceptance for diffractive protons with the $\beta^{*} = 1540$\,m optics
as a function of $t$ and $\xi \equiv \Delta p/p$ is 
shown in Figure~\ref{fig_txiaccept}. For $|t|>0.002$ all protons are detected
independent of $\xi$. In addition, there is acceptance for protons with 
$0.02\lesssim\xi\lesssim0.2$ independent of $t$.

Folding the acceptance with an assumed cross-section
$\frac{d\sigma}{dt\,d\xi} \propto \frac{1}{\xi} {\rm e}^{-7 |t|}$ yields a
total acceptance of more than 90\,\%. The $\xi$ 
resolution using the RP stations at 147\,m and 220\,m with the dipole D2 in
between is 0.5\,\%. For the optics of Scenario 3, an 
improvement of a factor 5 will be attempted.

Table~\ref{tab_diffracrates} shows the event rates of diffractive processes
for the running scenarios 1 and 2. 

\begin{table}[ht]  
\begin{center}
\scalebox{0.9}{  
\begin{tabular}{|c|c|c|c|c|c|c|}
\hline 
Process & Estimated cross-section & \multicolumn{2}{c|}{Rate at $\mathcal{L}\,[{\rm cm}^{-2}{\rm s}^{-1}] =$}\\
        &                         & $1.6 \times 10^{28}$ &  $2.4 \times 10^{29}$\\
\hline
Elastic scattering & 30\,mb       & 480\,Hz & 7.2\,kHz \\
Single diffraction & 14\,mb       & 240\,Hz & 3.6\,kHz \\
Double diffraction & 7\,mb        & 120\,Hz & 1.8\,kHz \\
Double Pomeron Exchange & 0.5\,mb &   8\,Hz & 120\,Hz  \\
\hline
\end{tabular}
}
\caption{Diffractive Event rates.}
\label{tab_diffracrates}  
\end{center}  
\end{table}
Some examples for exclusive production by DPE are listed in  
Table~\ref{tab_exclusiveDPE}
with their expected rates at three different luminosities.

\begin{table}[ht]  
\begin{center}
\scalebox{0.9}{  
\begin{tabular}{|c|c|c|c|c|c|c|}\hline 
X              & $\sigma$ & Decay channel & 
      BR            & \multicolumn{3}{c|}{Rate at $\mathcal{L}\,[{\rm cm}^{-2}{\rm s}^{-1}] =$}\\
               &          &               &    
                    & $2.4 \times 10^{29}$ & $10^{31}$ & $10^{33}$ \\
\hline\hline
excl. Dijets        & 7\,nb              & jj                                         &
 1                  & 720 / h            & 10 / s    & 1000 / s \\
($E_{T} > 10$\,GeV) &          &                                              &
                    &                    &           &         \\
\hline
$\chi_{c0}$    & 3\,$\mu$b & $\gamma J/\psi \rightarrow \gamma \mu^{+} \mu^{-}$&
 $6 \times 10^{-4}$ & 1.5 / h            & 62 / h    & 6200 / h  \\
(3.4\,GeV)     &           & $\pi^{+} \pi^{-} K^{+} K^{-}$ &
 0.018              & 46 / h             & 1900 / h  & 53 / s\\
\hline
$\chi_{b0}$    &  4\,nb   & $\gamma Y \rightarrow \gamma \mu^{+} \mu^{-}$&
 $\le 10^{-3}$          & $\le$ 0.07 / d      & $\le$3 / d     & $\le$ 300 / d \\
(9.9\,GeV)     &          &                                              &
                    &                    &           &         \\
\hline
H (SM)         & 3\,fb    & $b \bar{b}$                                  &
 0.68               & 0.02 / y           & 1 / y     & 100 / y \\
(120\,GeV)     &          &                                              &
                    &                    &           &         \\
\hline
\end{tabular}
}
\caption{Examples of exclusive DPE processes (p + p $\rightarrow$ p + X + p).
The rates do not account for any acceptance or analysis cuts. For 
cross-sections see e.g.~\protect\cite{kmr}.}
\label{tab_exclusiveDPE}  
\end{center}  
\end{table}
Evidently, detecting a 120\,GeV Higgs is hopeless at luminosities below
$10^{33}{\rm cm}^{-2}{\rm s}^{-1}$, but other interesting processes are
within reach even at low luminosity where the leading proton acceptance 
is very good.

At standard LHC running conditions ($\beta^{*} = 0.5\,$m) 
and with Roman Pots at 220\,m,
the lower edge of the $\xi$ acceptance of the Roman Pot trigger 
would be at 0.025 corresponding to a diffractive mass of 350\,GeV 
for exclusive production in the limit $\xi_{1} = \xi_{2}$.
Furthermore, the $\xi$ resolution is only 0.001, corresponding to 14\,GeV 
in the diffractive mass. Additional RP stations in the cryogenic LHC region
(at 308\,m, 338\,m and 420\,m) would bring the acceptance limit down to 
$\xi > 0.002$
and the resolution to $\sigma_{\xi} \gtrsim 0.0002$. Due to technical
difficulties such stations 
are not forseen in the current LHC design but may be considered at a later
stage after ruling out other possibilities (e.g. local optics modifications
for an enhanced dispersion at 220\,m).

\end{document}